\documentclass[preprint,12pt]{emulateapj-rtx4}
\usepackage[varg]{txfonts}
\usepackage{graphicx}
\usepackage{booktabs}
\usepackage{afterpage}
\usepackage{changepage}
\usepackage{chngcntr}
\bibpunct{(}{)}{;}{a}{}{,}
\usepackage{color}
\usepackage[caption=false]{subfig}
\usepackage[colorlinks=true,citecolor=blue,urlcolor=green]{hyperref}

\usepackage{longtable}

\newlength{\offsetpage}
{\end{adjustwidth}}

\begin{document}

\title{The K2-ESPRINT Project II: Spectroscopic follow-up of three exoplanet systems from Campaign 1 of K2$^{\star}$}

\hyphenation{Kepler}

\author{Vincent~Van~Eylen$^{1,2}$, Grzegorz Nowak$^{3,4}$, Simon~Albrecht$^1$, Enric Palle$^{3,4}$, Ignasi Ribas$^5$, Hans Bruntt$^1$, Manuel Perger$^5$, Davide Gandolfi$^{6,7}$, Teriyuki Hirano$^8$, Roberto Sanchis-Ojeda$^{9,10}$, Amanda Kiilerich$^1$, Jorge P. Arranz$^{3,4}$, Mariona Badenas$^{11}$, Fei Dai$^2$, Hans~J.~Deeg$^{3,4}$, Eike~W.~Guenther$^{12}$, Pilar Monta\~n\'es-Rodr\'iguez$^{3,4}$, Norio Narita$^{13,14,15}$, Leslie A.~Rogers$^{16}$, V\'ictor J. S. B\'ejar$^{3,4}$, Tushar S. Shrotriya$^1$, Joshua N. Winn$^2$, Daniel Sebastian$^{12}$}

\affil{$^1$ Stellar Astrophysics Centre, Department of Physics and Astronomy, Aarhus University, Ny Munkegade 120, \\
DK-8000 Aarhus C, Denmark}
\affil{$^2$ Department of Physics, and Kavli Institute for Astrophysics and Space Research, Massachusetts Institute of Technology, Cambridge, MA 02139, USA\label{inst2}}
\affil{$^3$ Instituto de Astrof\'\i sica de Canarias (IAC), 38205 La Laguna, Tenerife, Spain}
\affil{$^4$ Departamento de Astrof\'\i sica, Universidad de La Laguna (ULL), 38206 La Laguna, Tenerife, Spain}
\affil{$^5$ Institut de Ci\'encies de l'Espai (CSIC-IEEC), Carrer de Can Magrans, Campus UAB, 08193 Bellaterra, Spain}
\affil{$^6$ Dipartimento di Fisica, Universit\'a di Torino, via P. Giuria 1, I-10125, Torino, Italy}
\affil{$^7$ Landessternwarte K\"onigstuhl, Zentrum f\"ur Astronomie der Universit\"at Heidelberg, K\"onigstuhl 12, D-69117 Heidelberg, Germany}
\affil{$^8$ Department of Earth and Planetary Sciences, Tokyo Institute of Technology, 2-12-1 Ookayama, Meguro-ku, Tokyo 152-8551, Japan}
\affil{$^9$ Department of Astronomy, University of California, Berkeley, CA 94720}
\affil{$^{10}$ NASA Sagan Fellow}
\affil{$^{11}$ Department of Astronomy, Yale University, New Haven, CT 06511, USA}
\affil{$^{12}$ Th\"uringer Landessternwarte Tautenburg, 07778 Tautenburg, Germany}
\affil{$^{13}$ National Astronomical Observatory of Japan, 2-21-1 Osawa, Mitaka, Tokyo 181-8588, Japan}
\affil{$^{14}$ SOKENDAI (The Graduate University for Advanced Studies), 2-21-1 Osawa, Mitaka, Tokyo 181-8588, Japan}
\affil{$^{15}$ Astrobiology Center, National Institutes of Natural Sciences, 2-21-1 Osawa, Mitaka, Tokyo 181-8588, Japan}
\affil{$^{16}$ Department of Astronomy and Division of Geological and Planetary Sciences, California Institute of Technology, MC249-17, 1200 East California Boulevard, Pasadena, CA 91125, USA}

\altaffiltext{$\star$}{Based on observations made with the NOT telescope under programme ID. 50-022/51-503 and 50-213(CAT), the TNG telescope under programme ID. AOT30.13, OPT15A\_33, and CAT14B\_121 and ESOs $3.6$\,m telescope at the La~Silla~Paranal~Observatory under programme ID 095.C-0718(A). }

\email{vincent@phys.au.dk}

\shorttitle{Spectroscopic follow-up of K2 Campaign~1}
\shortauthors{Van Eylen et al.}

\received{receipt date}
\revised{revision date}
\accepted{acceptance date}

\begin{abstract}
{We report on Doppler observations of three transiting planet candidates that were detected during Campaign~1 of the K2 mission. The Doppler observations were conducted with FIES, HARPS-N and HARPS. {We measure the mass of K2-27b (EPIC 201546283b), and provide constraints and upper limits for EPIC 201295312b and EPIC 201577035b.} K2-27b is a warm Neptune orbiting its host star in 6.77 days and has a radius of $4.45^{+0.33}_{-0.33}~\mathrm{R_\oplus}$ and a mass of $29.1^{+7.5}_{-7.4}~\mathrm{M_\oplus}$, which leads to a mean density of $1.80^{+0.70}_{-0.55}~\mathrm{g~cm^{-3}}$. 
EPIC 201295312b is smaller than Neptune with an orbital period of 5.66 days, radius $2.75^{+0.24}_{-0.22}~\mathrm{R_\oplus}$ and we {constrain the mass to be below $12~\mathrm{M_\oplus}$ at 95\% confidence. We also find a long-term trend indicative of another body in the system.} 
EPIC 201577035b, {previously confirmed as the planet K2-10b}, is smaller than Neptune orbiting its host star in 19.3 days, with radius $3.84^{+0.35}_{-0.34}~\mathrm{R_\oplus}$. {We determine its mass to be $27^{+17}_{-16}~\mathrm{M_\oplus}$, with a 95\% confidence uppler limit at $57~\mathrm{M_\oplus}$, and mean density $2.6^{+2.1}_{-1.6}~{\rm g~cm}^{-3}$.} These measurements join the relatively small {collection of planets smaller than Neptune with measurements or constraints of the mean density}. Our code for performing K2 photometry and detecting planetary transits is now publicly available.
}
\end{abstract}


\maketitle

\section{Introduction}

Although data from the K2 mission \citep{howell2014} has only been available for six months, it has already led to several notable exoplanet discoveries. For example, a sub-Neptune orbiting a bright star \citep[using only the 9 days of Engineering Test Data][]{vanderburg2015}, three super-Earths orbiting a bright M dwarf star \citep{crossfield2015}, a disintegrating rocky planet with a cometary head and tail \citep{sanchisojeda2015}, two super-Earth planets orbiting a nearby cool star \citep{petigura2015} and two additional planets orbiting the known hot Jupiter host star WASP-47 \citep{becker2015}. Based on the Campaign~1 photometry, the first lists of planetary candidates have been produced \citep{foremanmackey2015,montet2015}.

As part of the \textit{Equipo de Seguimiento de Planetas Rocosos INterpretando sus Tr\'ansitos} (ESPRINT) project \citep[see][]{sanchisojeda2015}, we present our radial velocity follow-up measurements of three Campaign~1 planet candidates (EPIC 201295312, EPIC 201546283, and EPIC 201577035), making use of the FIES \citep{telting2014}, HARPS-N \citep{cosentino2012} and HARPS \citep{mayor2003} spectrographs. Measurements of masses for planets smaller than $4-5~\mathrm{R}_\oplus$ are notoriously difficult \citep{marcy2014}, but are of importance to constrain interior models for sub-Neptune planets \citep[e.g.][]{rogers2015}.

In the next section we present our analysis pipeline for K2 photometry, including aperture photometry, light curve detrending and planet search algorithms (the Python code used for the analysis is publicly available on GitHub\footnote{https://github.com/vincentvaneylen}). We describe the planet characterization via spectroscopic observations in Section~\ref{sec:spectroscopy}, and discuss the results in Section~\ref{sec:discussion}.

\begin{figure*}[!htbp]
\centering
\resizebox{0.33\hsize}{0.25\hsize}{\includegraphics{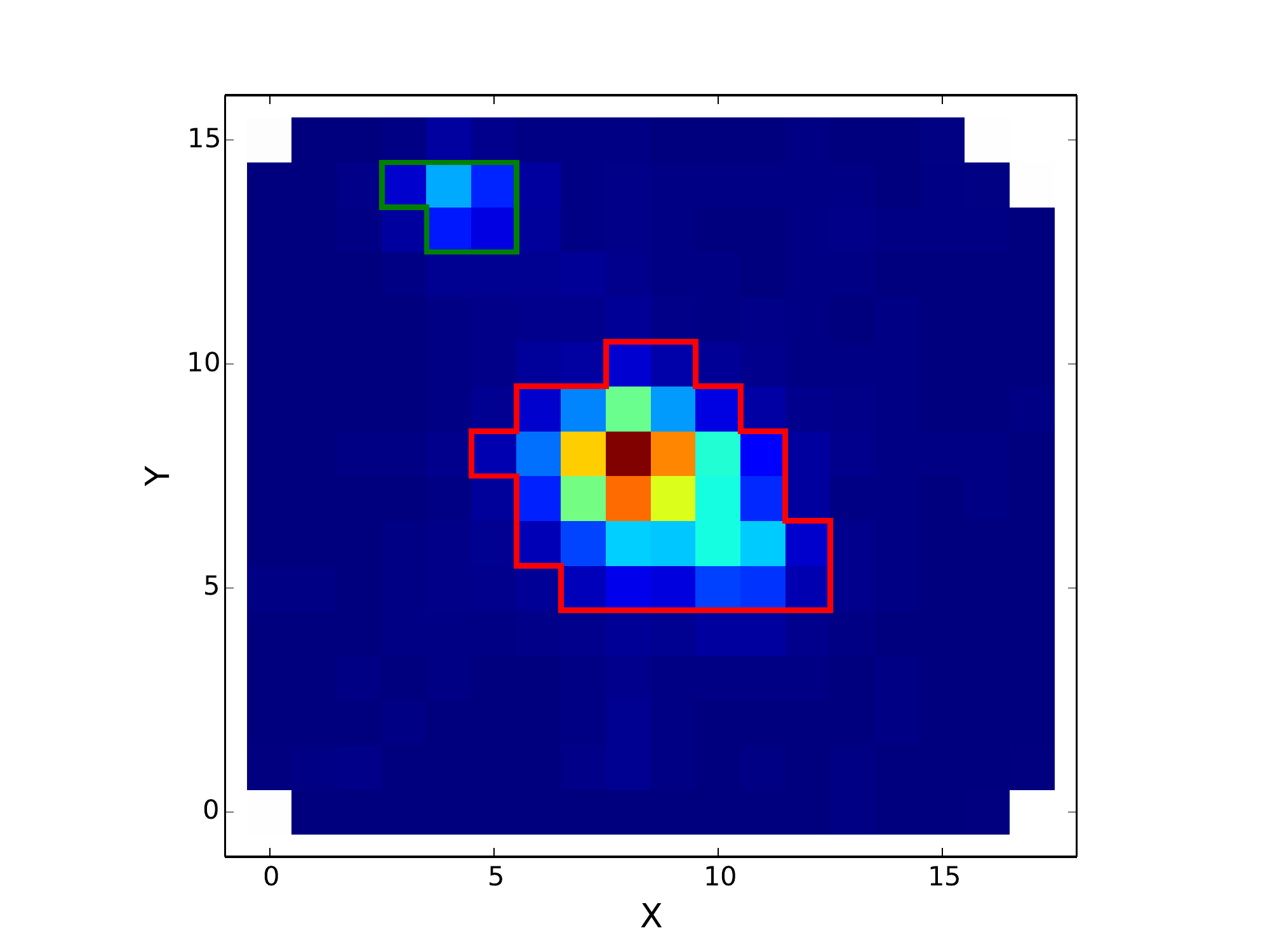}}
\resizebox{0.33\hsize}{0.25\hsize}{\includegraphics{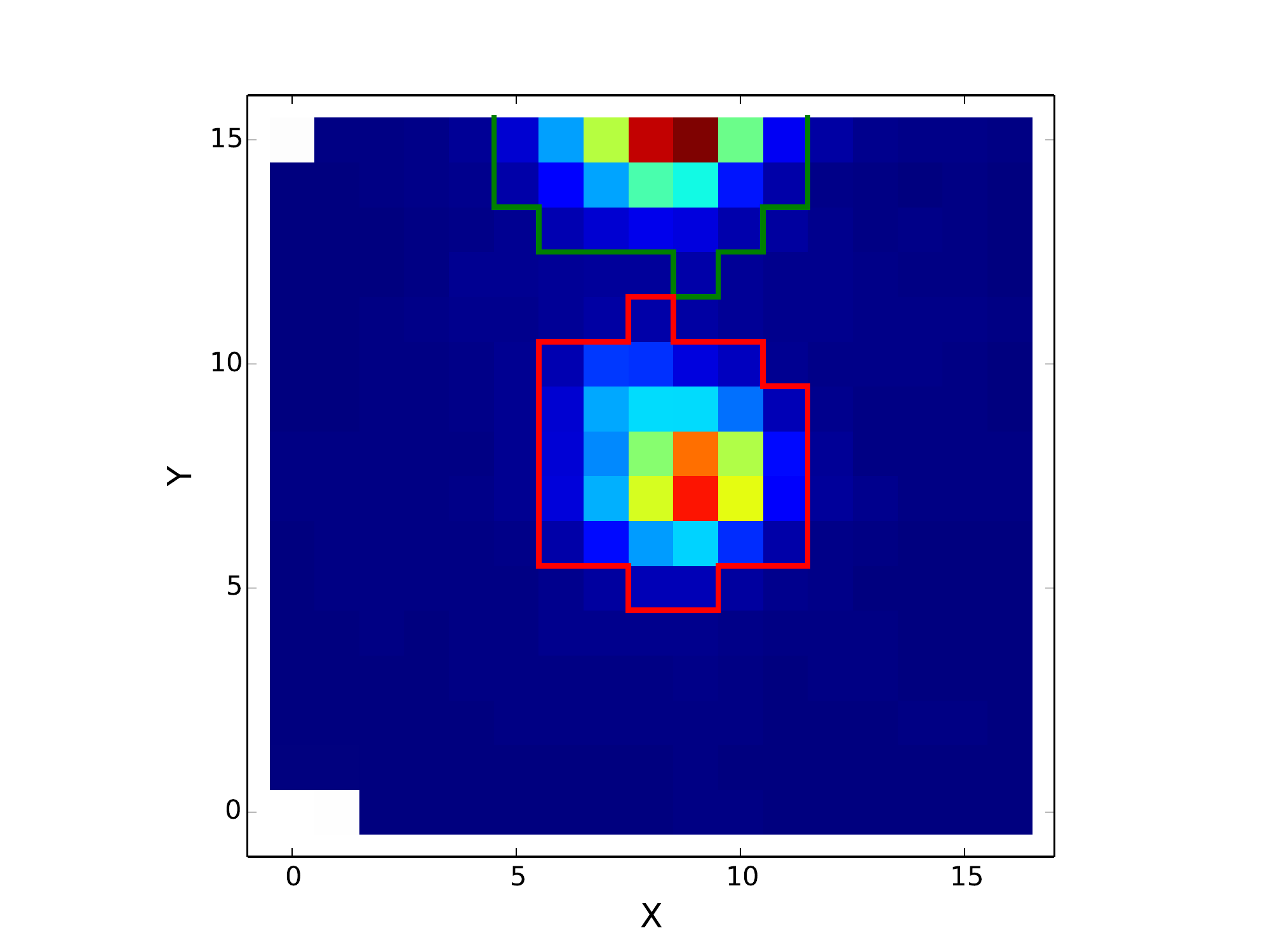}}
\resizebox{0.33\hsize}{0.231\hsize}{\includegraphics{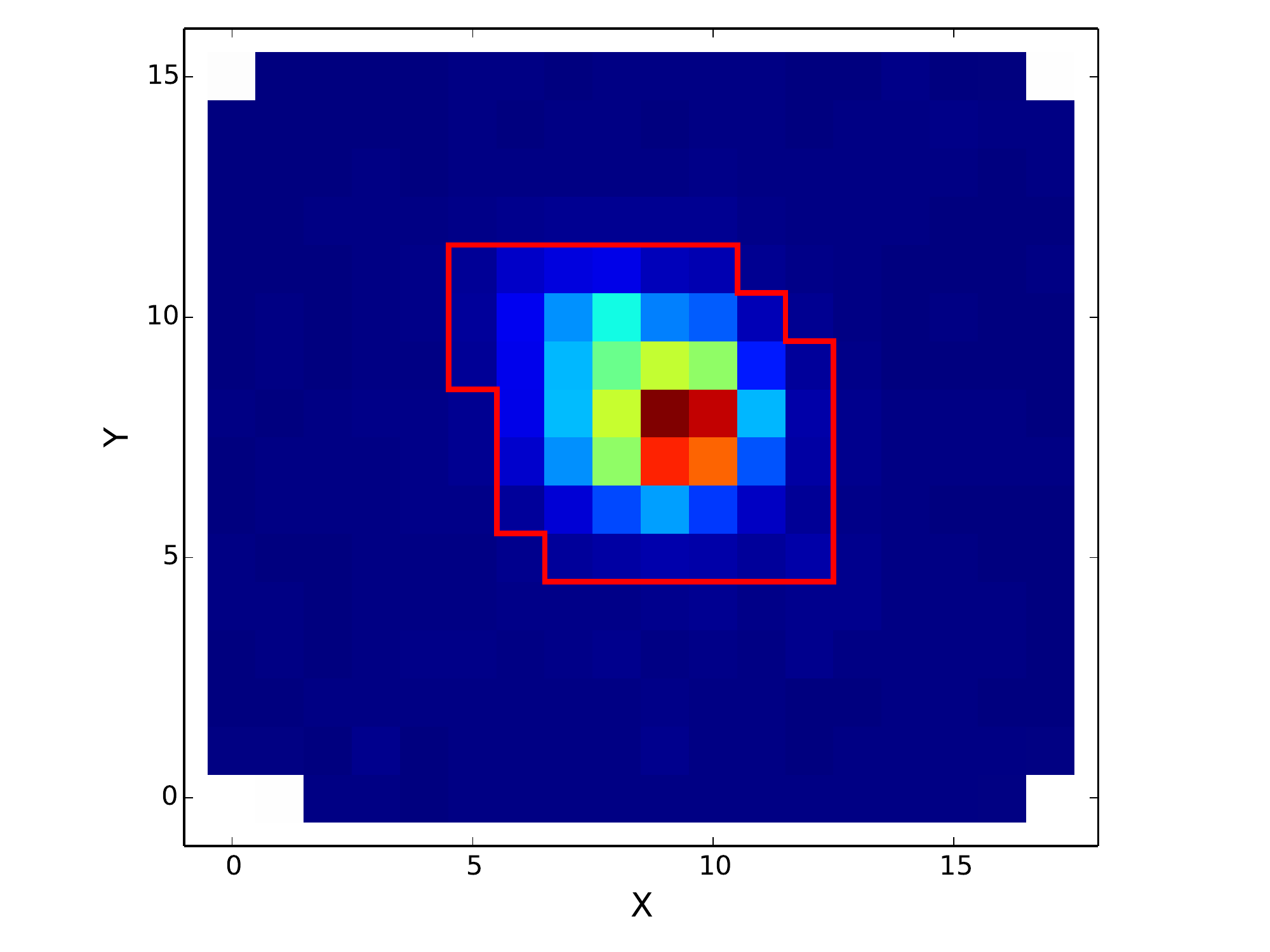}}
\caption{Pixel masks for EPIC 201295312 (left), EPIC 201546283 (middle) and EPIC 201577035 (right). The colors indicate the electron count, going from red (high) to low (blue). Pixel masks above a threshold electron count are encircled. We use red for those pixels included in the light curve and green for those assumed to be caused by other stars.\label{fig:aperture}}
\end{figure*}

\section{Photometry}
\label{sec:photometry}

Unlike for the primary \textit{Kepler} mission, K2 photometry is primarily delivered in the form of pixel files without mission-defined aperture masks, and the task of finding planet candidates rests upon the community rather than upon the mission team. Because the mission operates with only two functioning reaction wheels \citep{howell2014} the pointing stability is more limited, which affects the photometric precision. Correction methods make use of the center-of-light offset \citep{vanderburg2014,lund2015} or use trends which are common to many stars \citep{foremanmackey2015,angus2015}.

We have developed a photometry pipeline consisting of the following independent modules: (1) Extract aperture photometry; (2) Perform light curve detrending; and (3) Search for transits and perform time-domain transit modeling. We describe these steps in the next sections.

Our analysis starts from K2 pixel mask files which can be downloaded from the MAST archive\footnote{See https://archive.stsci.edu/k2/data\_search/search.php}. We perform simple aperture photometry on these pixel masks. First we sum the flux per pixel over the full time series of the K2 campaign. Subsequently, the median flux over the different pixels is calculated. As long as the field is not too crowded the median flux is a fairly good estimate of the background flux. The stellar flux can then be identified as the flux that exceeds the background flux by some predetermined threshold (typically $1.05 \times \textrm{median}$). We include the pixels exceeding the threshold and group them according to whether they are spatially adjacent. If two or more spatially disjoint groups are detected, the largest pixel group is selected (the other, smaller groups are assumed to be caused by other stars and are ignored). The results are shown in Figure~\ref{fig:aperture} for the three stars discussed in this work. Once
the aperture is selected in this way, for each time step the total flux is calculated by summing the flux of all pixels in the aperture, and the flux centroid position is calculated based on the flux-weighted mean $X$ and $Y$ coordinates of the group of pixels. We also subtract the background flux, which is estimated as the median of the pixels after iteratively clipping all $3\sigma$ outliers.

For the light curve detrending, we use a linear fit to the centroid positions, a technique first used successfully for \textit{Spitzer} transit observations \citep[e.g.][]{desert2009,vaneylen2014,desert2015}, and which is similar to techniques developed for K2 photometry by others \citep{vanderburg2014,lund2015,sanchisojeda2015}. In summary, the light curve is divided into chunks of specified length (typically 300 data points), and a polynomial function of centroid position and time is fitted to the flux in each chunk. More precisely, for centroid coordinates $X_c$ and $Y_c$ (calculated relative to the mean centroid position), time $T$, and flux $F$, we fit the model $M$:

\vspace{-1em}
\begin{eqnarray*}
   M = t_0 + t_1 T + t_2 T^{2} + t_3 T^3 + x_1 X_c + x_2 X_c^2 + x_3 X_c^3\\
   + y_1 Y_c + y_2 Y_c^2 + y_3 Y_c^3 + z_1 X_c Y_c,
\end{eqnarray*}

where $t_i$, $x_i$, $y_i$ and $z_1$ are fitting parameters. For each chunk, the light curve flux is then divided by the model to remove variability caused by spacecraft pointing variations (which cause flux variations due to different pixel sensitivities) as well as long-term astrophysical variations. We have compared this technique with the ones employed by \cite{sanchisojeda2015} and \cite{vanderburg2014}, and found the light curve quality and the transit parameters to be similar.

\begin{figure*}
\centering
\resizebox{0.3\hsize}{!}{\includegraphics{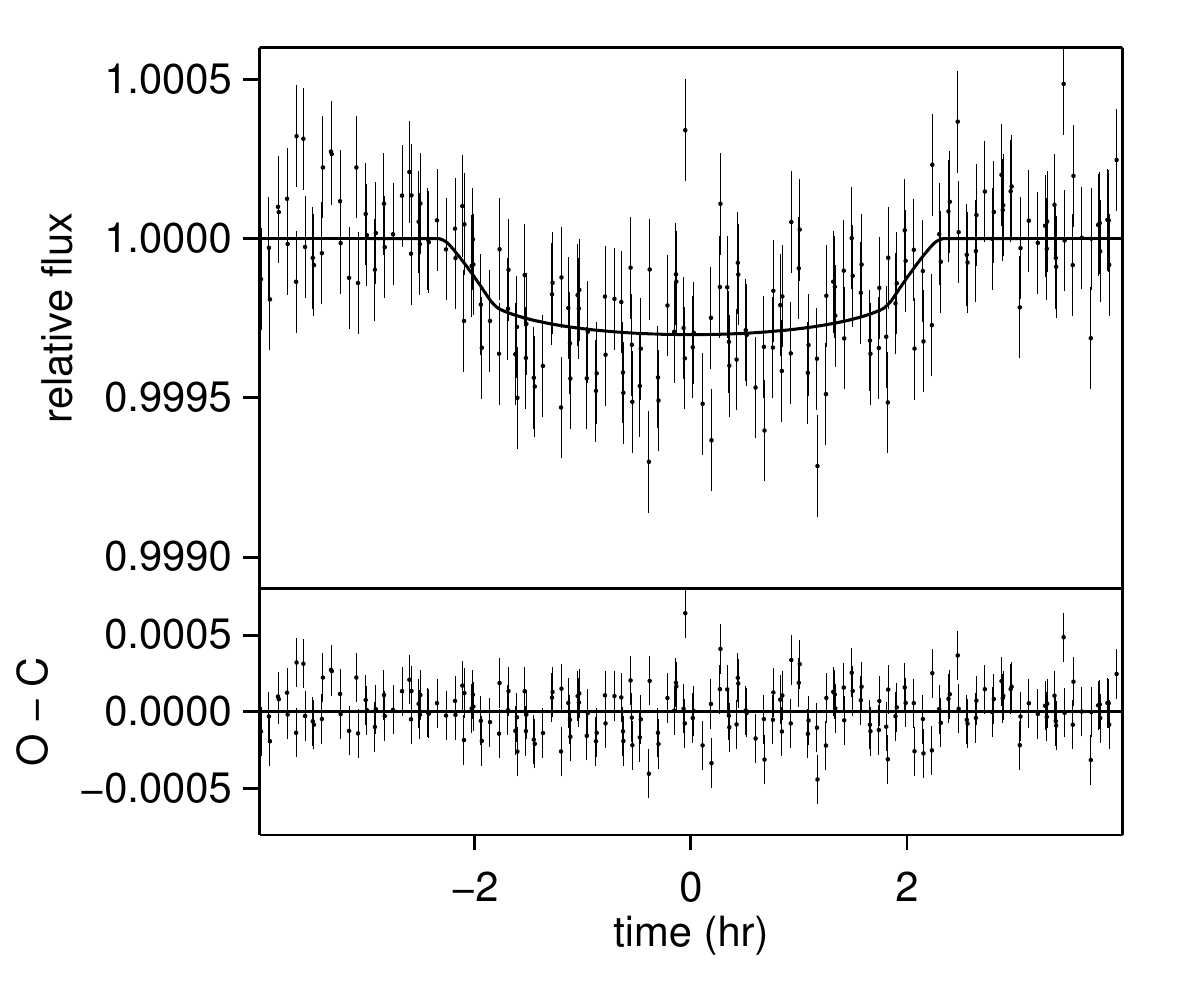}}
\resizebox{0.3\hsize}{!}{\includegraphics{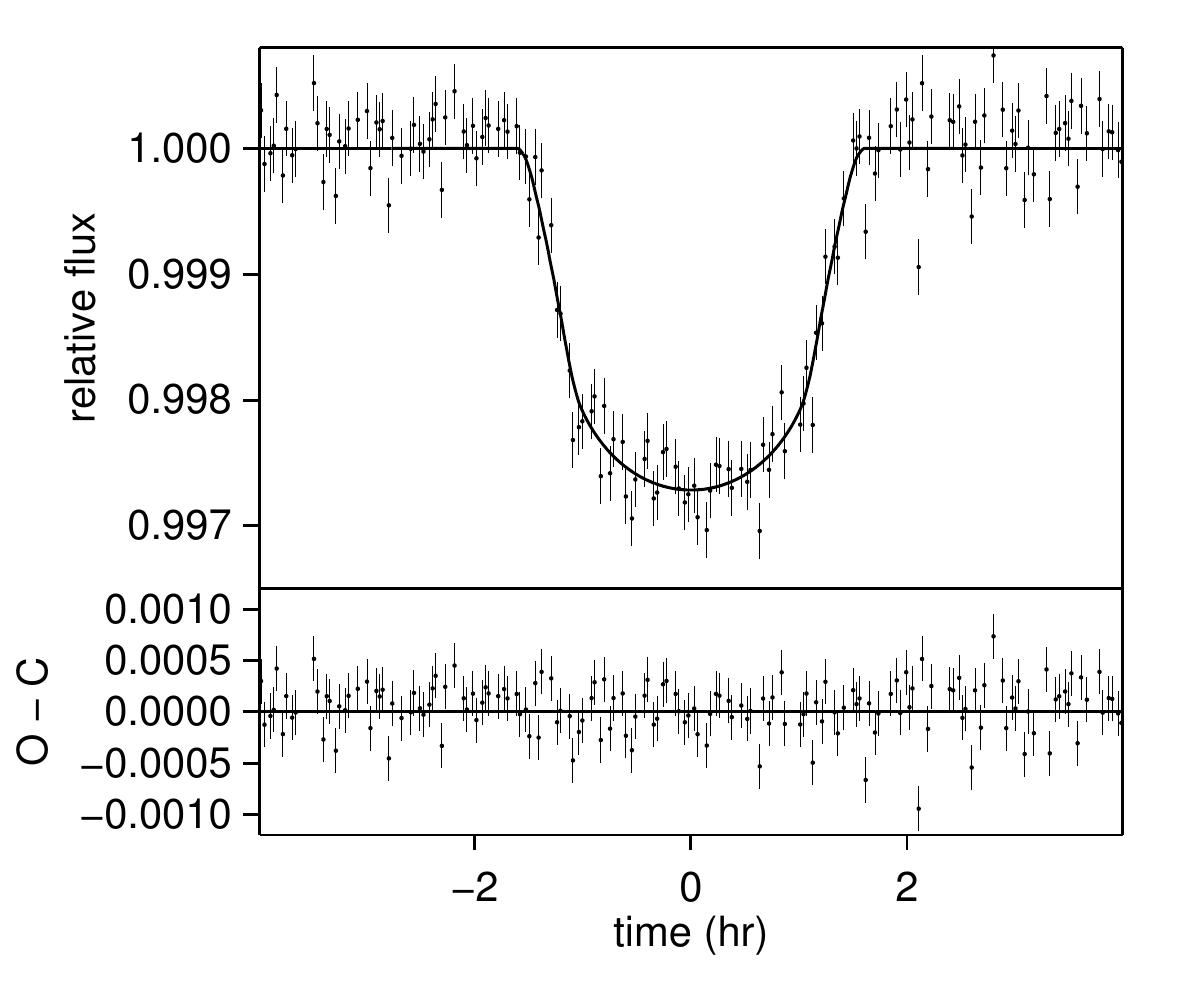}}
\resizebox{0.3\hsize}{!}{\includegraphics{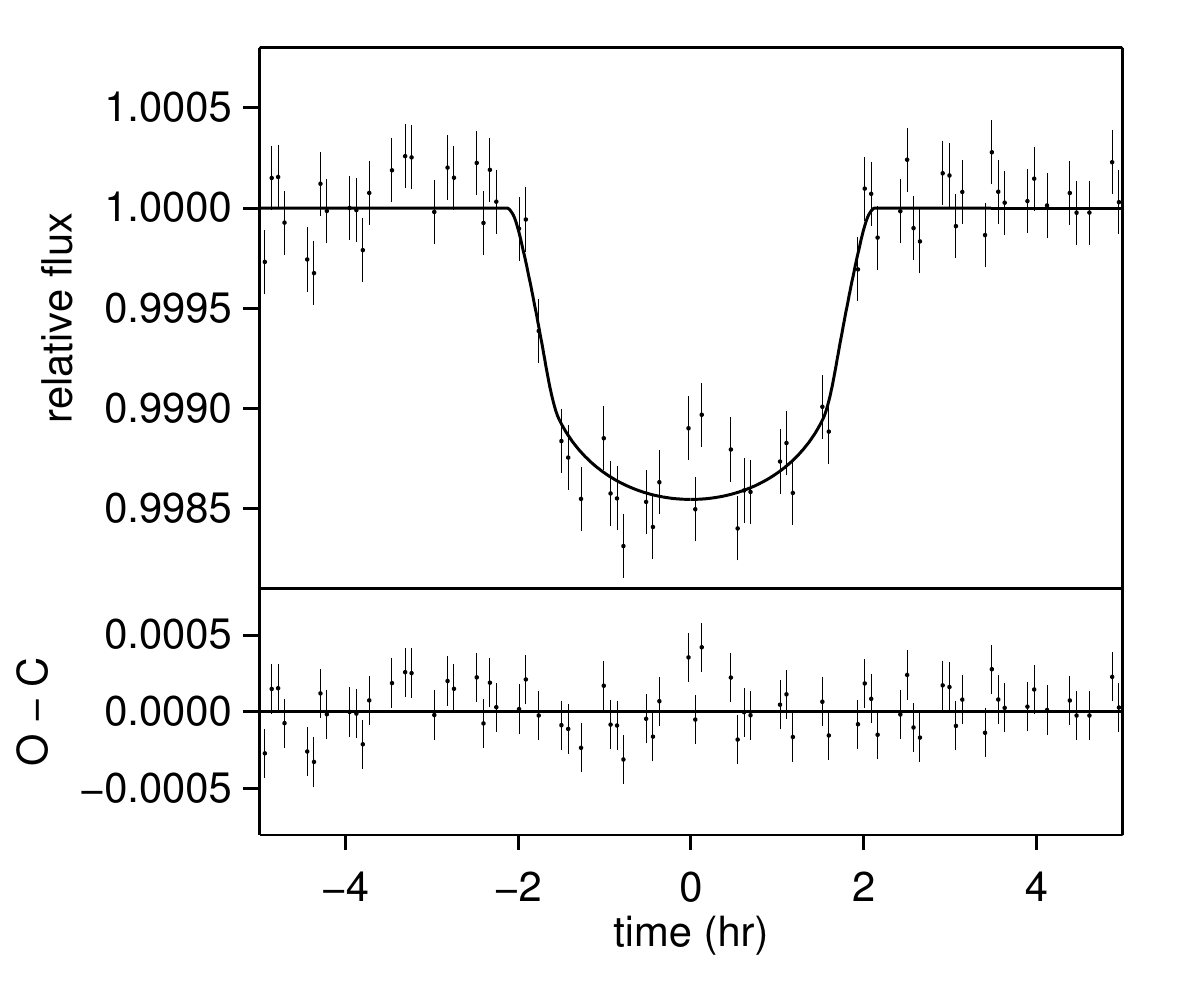}}
\caption{K2 reduced photometry folded by the best period for EPIC 201295312 (left), EPIC 201546283 (middle) and EPIC 201577035 (right). The best fitted model is shown with a solid line as well as the residuals after subtracting the model.\label{fig:transits}}
\end{figure*}

To search for transit events we subsequently run a ``box least square'' search on the light curves \citep{kovacs2002}, which detects periodic transit-like events\footnote{We used an implementation of this algorithm in Python by Ruth Angus and Dan Foreman-Mackey; see https://github.com/dfm/python-bls.}. These are then visually inspected in order to check if they are indeed transit-shaped.

Based on an initial analysis of the light curves and ground based imaging, interesting planets were selected for spectroscopic follow-up. Bright planet candidates were observed using the FastCam (FC) lucky imaging camera at the 1.5-m Carlos Sanchez Telescope (TCS) in Tenerife. All images were bias subtracted and then shifted and co-added using FC specific software to produce a final, high SNR, high resolution image. Objects that appeared to be isolated, were then moved forward in the confirmation process to be observed with FIES. We obtained observations (45-60 minute exposures) with the FIES spectrograph for a first detailed stellar characterization and a small number of Radial Velocities (RVs). Systems which show RV scatter less than 20\,m\,s$^{-1}$ {are selected for further observations}. For Campaign~1, these efforts were focused on EPIC~201295312, EPIC~201546283, and EPIC~201577035.

\section{Ground based follow-up observations}
\label{sec:spectroscopy}

These systems were recently discussed by \cite{montet2015}. EPIC~201577035 was validated as a genuine planet (also called K2-10b in the NASA Exoplanet Archive\footnote{http://exoplanetarchive.ipac.caltech.edu/}), but high-resolution Adaptive Optics images conducted by these authors revealed faint stellar companions for EPIC~201295312 and EPIC~201546283, at distances of 3 and 8 arcsec respectively. This complicates the planet validation because K2 apertures span many pixels (see Figure~\ref{fig:aperture}; each pixel measures $3.98 \times 3.98$ arcsec). Therefore it can be difficult to be certain that the target star is truly the host of the transiting planet candidate. Consequentially, \cite{montet2015} were unable to validate the planetary candidates (despite assigning false positive probabilities below $10^{-4}$ in both cases). We measured RVs using the standard data reduction pipelines for the HARPS and HARPS-N spectrographs. For the case of FIES we used the approach described in \cite{gandolfi2015}.

To derive stellar parameters, to measure stellar reflex motion and finally to determine the planetary mass we observed these three systems throughout Spring 2015 with the HARPS-N spectrograph mounted on the TNG on La\,Palma and the HARPS spectrograph on ESO's 3.6m telescope at La\,Silla. The exposure times varied between 15\,min and 45\,min for the different systems and instruments, and we used standard setups. All RVs for the three systems are available in electronic form from the ApJ webpage.

\section{Parameter estimation}

\subsection{Estimation of stellar parameters}
\label{sec:stellar}

{Before modeling the RVs we co-added the available spectra for each system to derive the stellar atmospheric parameters using the VWA software\footnote{{\url https://sites.google.com/site/vikingpowersoftware/home}} developed by \cite{bruntt2012}. This approach includes systematic errors in the uncertainty estimate. For the signal-to-noise ratio in our combined spectra, the uncertainty in the stellar parameters is dominated by this systematic noise floor rather than by photon noise, so that the uncertainties in parameters for different stars are sometimes very similar.
With obtained values of the effective stellar temperature ($T_{\rm eff}$), the stellar surface gravity ($\log g$), and metallicity ([Fe/H]) as input, we then used BaSTI evolution tracks\footnote{{\url http://albione.oa-teramo.inaf.it/} } to infer the stellar mass, radius, and obtain an age estimate. Here we used the SHOTGUN method \citep{stello2009}.}

{
In parallel we also obtained spectra with the High Dispersion spectrograph (HDS) mounted to the Subaru telescope, one spectrum for each system. These spectra have been analyzed 
following \cite{takeda2002} \citep[see also][]{hirano2012}. There is agreement between the set of parameters obtained with the two different data sets and methods, with one important exception, i.e.\ the stellar radius of EPIC 201295312 (the HDS radius is $1.91 \pm 0.11~R_\odot$, the VWA radius is $1.52 \pm 0.10~R_\odot$). This is an important parameter, because any uncertainty or error in this parameter translates directly into the planetary radius and the planetary density. However, we have not been able to track down the cause for this disagreement. Here we just note that the two methods agree well on all other stellar parameters and for the other systems and that for evolved stars (such as EPIC 201295312), different isochrones can lay close to each other, complicating the stellar characterization. Because the HDS spectrum has a 
lower signal-to-noise ratio than the combined HARPS-N spectrum, we adopt the VWA values in the analysis.}

\subsection{Estimation of orbital and planetary parameters}
\label{sec:orbital}

\paragraph{Photometric model}

{We modeled the K2 light curves together with the RVs obtained from the FIES, HARPS-N, and HARPS spectrographs. We use the transit model by \cite{mandel2002} and a simple Keplerian RV model. The transit model used was binned to 30 minutes, to match the integration time of the \textit{Kepler} observations. The model parameters mainly constrained by photometry (see Figure~\ref{fig:transits}) include the orbital Period ($P$), a particular time of mid-transit ($T_{\rm min}$), the stellar radius in units of the orbital semi-major axis ($R_{\star}/a$), the planetary radius in units of the stellar radius ($R_{\rm p}/R_{\star}$), and the cosine of the orbital inclination ($\cos i_{\rm o}$). We further assumed a quadratic limb-darkening law (with two parameters, $u_1$ and $u_2$).}

\paragraph{RV model}
{The Keplerian RV model introduces additional parameters, i.e.\ the semi-amplitude of the projected stellar reflex motion ($K_{\star}$), systemic velocities for each spectrograph ($\gamma_{\rm spec}$), the orbital eccentricity ($e$), and the argument of periastron ($\omega$). For our analysis, we use $\sqrt{e} \cos\omega$ and $\sqrt{e} \sin\omega$ to avoid boundary issues \citep[see e.g.][]{lucy1971}. For one system (EPIC\,201295312) we found evidence for a long term drift which we model with a second order polynomial function of time.}

\paragraph{Prior information}
{For all parameters in all three systems we use a flat prior if not mentioned otherwise in this paragraph. 
We use priors on the quadratic limb darkening parameters $u_1$ and $u_2$, selected from the tables provided by \cite{claret2013} appropriate for the Kepler bandpass. In particular we placed a Gaussian prior on $u_1+u_2$ with a width of $0.1$. The difference $u_1-u_2$ was held fixed at the tabulated value, since this combination is weakly constrained by the data. The stellar density obtained from the analysis of the stellar spectra (Section~\ref{sec:stellar}) is used as a Gaussian prior in our fit to the photometric and RV data. This impacts the confidence intervals we obtain for $e$ and $\omega$ \cite[e.g.][]{vaneylen2015}. We further assume an eccentricity prior $\frac{dN}{de} \propto \frac{1}{(1+e)^4} - \frac{e}{2^4}$ as obtained by \cite{shen2008}, and require that the planet and star do not have crossing orbits.}

{For EPIC\,201295312 we found indications of a long term drift, using the FIES and HARPS-N data sets. Unfortunately the HARPS data points have been taken after the FIES and HARPS-N campaigns were finished. This complicates the characterization of the long term trend, as potential offsets in the RV zero points of the spectrographs could lead to biases. However, previous studies, such as \cite{lopezmorales2014} (55\,Cnc) and \cite{desidera2013} (HIP\,11952) found that the RVs of HARPS and HARPS-N agree within their uncertainties. In this study, we find the same for EPIC\,201546283 (see below). Therefore, we proceed and impose a Gaussian prior with zero mean and a $\sigma$ of $5$~m\,s$^{-1}$ on the difference in the systemic velocity of these two spectrographs for EPIC\,201295312.
We also assumed the period to be constant in all three systems as we could detect no sign of {significant} Transit Timing Variations.}

\paragraph{Parameter estimation}
{To estimate the uncertainty intervals for the parameters we use a Markov Chain Monte Carlo (MCMC) approach \citep[see, e.g.,][]{tegmark2004}. Before starting the MCMC chain we added ``stellar jitter'' to the internally estimated uncertainties for FIES, HARPS-N, and HARPS observations, so that the minimum reduced $\chi^2$ for each dataset alone is close to unity.}

{For each system we run three simple chains with $10^6$ steps each, using the Metropolis-Hastings sampling algorithm. The step size was adjusted to obtain a success rate of $\approx 0.25$. We removed the first $10^4$ points from each chain and checked for convergence via visual inspection of trace plots and employing the Gelman and Rubin Diagnostic \citep{gelman1992}. Here we find that for all parameters for all systems $R < 1.01$. The uncertainty intervals presented below have been obtained from the merged chains. In Table \ref{tab:planet_para} we report the results derived from the posterior, quoting uncertainties excluding $15.85$\,\% of all values at both extremes, encompassing $68.3$\,\% of the total probability. The key result is $K_\star$, which together with the orbital period, inclination, and stellar mass determines planetary mass and bulk density.}

\begin{table*}
  \begin{center}
    \caption{System parameters \label{tab:planet_para}}
    \smallskip
    \begin{tabular}{l  r@{$\pm$}l r@{$\pm$}l r@{$\pm$}l   }
      \tableline\tableline
      \noalign{\smallskip}
      Parameter &  \multicolumn{2}{c}{EPIC  201295312   }  &    \multicolumn{2}{c}{EPIC 201546283 } &    \multicolumn{2}{c}{EPIC 201577035} \\
      \noalign{\smallskip}
      \hline
      \noalign{\smallskip}
      \multicolumn{5}{c}{Basic properties} \\
      \noalign{\smallskip}
      \hline
      \noalign{\smallskip}
      {2MASS ID}	 				& \multicolumn{2}{c}{11360278-0231150} 	&    \multicolumn{2}{c}{11260363+0113505} 	&    \multicolumn{2}{c}{11282927+0141264} \\
      {Right Ascension}				& \multicolumn{2}{c}{11 36 02.790}  	&    \multicolumn{2}{c}{11 26 03.638} 		&    \multicolumn{2}{c}{11 28 29.269} \\
      {Declination} 				& \multicolumn{2}{c}{-02 31 15.17}  	&    \multicolumn{2}{c}{+01 13 50.66} 		&    \multicolumn{2}{c}{+01 41 26.29} \\
      {Magnitude (\textit{Kepler})}		& \multicolumn{2}{c}{12.13}  		&    \multicolumn{2}{c}{12.43} 			&    \multicolumn{2}{c}{12.30} \\
      \hline
      \noalign{\smallskip}
      \multicolumn{5}{c}{Stellar parameters from spectroscopy} \\
      \noalign{\smallskip}
      \hline
      \noalign{\smallskip}
      Effective Temperature, $T_{\rm_{eff}}$ (K) 			& $5830$ & $70$ 		& $5320$ & $70$ 		& $5620$ & $70$ \\
      Surface gravity, $\log g$ (cgs)            			& $4.04$ & $0.08$ 		& $4.60$ & $0.08$ 		& $4.50$ & $0.08$ \\
      Metallicity, [Fe/H]                        			& $0.13$ & $0.07$ 		& $0.14$ & $0.07$ 		& $-0.07$ & $0.07$ \\
      Microturbulence (km\,s$^{-1}$)             			& $1.2$ & $0.07$ 		& $0.8$ & $0.07$ 		& $0.9$ & $0.07$ \\
      Projected rotation speed, $v \sin i_{\star}$ (km\,s$^{-1}$) 	& $5$ & $1$ 			& $1$ & $1$ 			& $3$ & $1$ \\
      Assumed Macroturbulence, $\zeta$ (km\,s$^{-1}$) 							& \multicolumn{6}{c}{$2$}  \\
      Stellar Mass,   $M_{\rm p} $ ($M_{\odot}$)			& $1.13$ & $0.07$ 		& $0.89$ & $0.05$ 		& $0.92$ & $0.05$ \\
      Stellar Radius, $R_{\rm p} $ ($R_{\odot}$)	 		& $1.52$ & $0.10$ 		& $0.85$ & $0.06$ 		& $0.98$ & $0.08$ \\
      Stellar Density, $\rho_\star$ (g cm$^{-3}$)$^{\rm a}$ 		& $0.45$ & $0.14$ 		& $2.04$ & $0.38$ 		& $1.38$ & $0.34$ \\
      \hline
      \noalign{\smallskip}
      \multicolumn{5}{c}{Fitting (prior) parameters} \\
      \noalign{\smallskip}
      \hline
      \noalign{\smallskip}
      Limb darkening prior $u_1 + u_2$ 					& $0.6752$ & $0.1$ 		& $0.7009$ & $0.1$ 		& $0.6876$ & $0.1$ \\
      Stellar jitter term HARPS (m\,s$^{-1}$) 				& \multicolumn{2}{c}{10.5}  	&    \multicolumn{2}{c}{6} 	&    \multicolumn{2}{c}{---} \\
      Stellar jitter term HARPS-N (m\,s$^{-1}$) 			& \multicolumn{2}{c}{6.5}  	&    \multicolumn{2}{c}{6} 	&    \multicolumn{2}{c}{7} \\
      Stellar jitter term FIES (m\,s$^{-1}$) 				& \multicolumn{2}{c}{20}  	&    \multicolumn{2}{c}{30} 	&    \multicolumn{2}{c}{5} \\
      \hline
      \noalign{\smallskip}
      \multicolumn{5}{c}{Adjusted Parameters from RV and transit fit} \\
      \noalign{\smallskip}
      \hline
      \noalign{\smallskip}
      Orbital Period, $P$ (days) 					& $5.65639$ & $0.00075$ 			& $6.77145$ & $ 0.00013$ 				& $19.3044$ & $0.0012$ \\
      Time of mid-transit, $T_{\rm min}$ (BJD$-2450000$) 		& $6811.7191$ & $ 0.0049$  			& $6812.8451$ & $0.0010$ 				& $6819.5814$ & $ 0.0021$ \\[3pt]
      Orbital Eccentricity, $e$ 					& \multicolumn{2}{c}{$0.12^{+0.22}_{-0.09}$} 	& \multicolumn{2}{c}{$0.16^{+0.10}_{-0.09}$} 		& \multicolumn{2}{c}{$0.31^{+0.16}_{-0.18}$} \\[3pt]
      Cosine orbital inclination, $\cos i_{\rm o}$ 			& \multicolumn{2}{c}{0.052$^{+0.039}_{-0.032}$}	& \multicolumn{2}{c}{$0.020^{+0.013}_{-0.013}$} 	& \multicolumn{2}{c}{$0.018^{+0.01}_{-0.012}$} \\[3pt]
      Scaled Stellar Radius, $R_\star/a$ 				& \multicolumn{2}{c}{$0.115^{+0.022}_{-0.011}$}	& \multicolumn{2}{c}{$0.0605^{+0.0051}_{-0.0038}$} 	& \multicolumn{2}{c}{$0.0349^{+0.0042}_{-0.0029}$} \\[3pt]
      Fractional Planetary Radius, $R_{\rm p}/R_\star$ 			& \multicolumn{2}{c}{$0.01654^{+0.00093}_{-0.00071}$} & \multicolumn{2}{c}{$0.0478^{+0.0013}_{-0.0008}$}& \multicolumn{2}{c}{$0.03570^{+0.0017}_{-0.0009}$} \\ [3pt]
      Linear combination limb darkening parameters (prior \& transit fit), $u_1+ u_2$, & $0.668$ & $0.098$ 		& $0.676$ & $0.082$ 					& $0.622$ & $0.092$ \\ [3pt]
      Stellar Density (prior \& transit fit), $\rho_\star$ (g\,cm$^{-3}$) & \multicolumn{2}{c}{$0.39^{+0.14}_{-0.16}$}	& \multicolumn{2}{c}{$1.80^{+0.70}_{-0.55}$}		& \multicolumn{2}{c}{$1.19^{+0.35}_{-0.34}$} \\ [3pt]
      Stellar radial velocity amplitude, $K_\star$ (m\,s$^{-1}$)      	& \multicolumn{2}{c}{$-0.18^{+2.6}_{-2.6}$} & \multicolumn{2}{c}{$10.8^{+2.7}_{-2.7}$} 		& \multicolumn{2}{c}{$7.3^{+4.6}_{-4.2}$} \\ [3pt]
      Linear RV term, $\phi_{1}$ (m\,s$^{-1}$/day) 			& $-0.03$ & $0.12$  & \multicolumn{2}{c}{---} & \multicolumn{2}{c}{---} \\
      Quadratic RV term, $\phi_{2}$ (m\,s$^{-1}$/day) 			& $0.0183$ & $0.0016$  & \multicolumn{2}{c}{---} & \multicolumn{2}{c}{---} \\
      Systemic velocity HARPS-N, $\gamma_{\rm HARPS-N}$ (km\,s$^{-1}$) 	& $44.561$ & $0.002$  				& $-37.772$ & $0.002$ 					& $8.203$ & $0.003$ \\
      Systemic velocity HARPS,   $\gamma_{\rm HARPS}$ (km\,s$^{-1}$)   	& $44.560$ & $0.006$ 				& $-37.776$ & $0.003$ 					& \multicolumn{2}{c}{---} \\
      Systemic velocity FIES,   $\gamma_{\rm FIES}$ (km\,s$^{-1}$)    	& $44.509$ & $0.010$ 				& $-37.979$ & $0.014$ 					& $ 8.062$ & $0.005$ \\
\noalign{\smallskip}
      \hline
      \noalign{\smallskip}
      \multicolumn{5}{c}{Indirectly Derived Parameters} \\
      \noalign{\smallskip}
      \hline
      \noalign{\smallskip}
      Impact parameter, b                           			& \multicolumn{2}{c}{$0.43^{+0.25}_{-0.26}$}	& \multicolumn{2}{c}{$0.30^{+0.21}_{-0.20}$} 		& \multicolumn{2}{c}{$0.40^{+0.27}_{-0.27}$} \\ [3pt]
      Planetary Mass,   $M_{\rm p} $ ($M_{\oplus}$)$^{\rm b}$ 		& \multicolumn{2}{c}{} 	& \multicolumn{2}{c}{$29.1^{+7.5}_{-7.4}$} 		& \multicolumn{2}{c}{$27^{+17}_{-16}$} \\ [3pt]
      {Mass upper limit (95\% confidence)}, $M_{\rm p} $ ($M_{\oplus}$)$^{\rm b}$ & \multicolumn{2}{c}{$12.0$} 		& \multicolumn{2}{c}{$41.5$} 				& \multicolumn{2}{c}{$57$} \\ [3pt]
      Planetary Radius, $R_{\rm p} $ ($R_{\oplus}$)$^{\rm b}$ 		& \multicolumn{2}{c}{$2.75^{+0.24}_{-0.22}$} 	& \multicolumn{2}{c}{$4.45^{+0.33}_{-0.33}$} 		& \multicolumn{2}{c}{$3.84^{+0.35}_{-0.34}$} \\[3pt]
      Planetary Density, $\rho_{\rm p}$ (g\,cm$^{-3}$) 			& \multicolumn{2}{c}{$\leq 3.3~(95\%$ confidence)} 	& \multicolumn{2}{c}{$1.80^{+0.70}_{-0.55}$} 		& \multicolumn{2}{c}{$2.6^{+2.1}_{-1.6}$} \\
      \noalign{\smallskip}
      \tableline
      \noalign{\smallskip}
      \noalign{\smallskip}
      \multicolumn{5}{l}{{\sc Notes} ---}\\
      \multicolumn{5}{l}{$^{\rm a}$ This value is used as a prior during the fitting procedure.}\\
      \multicolumn{5}{l}{$^{\rm b}$ Adopting an Earth radius of $6371$~km and mass of $5.9736\cdot10^{24}$~kg.}\\
      \noalign{\smallskip}

    \end{tabular}
  \end{center}
\end{table*}

\section{Results}
\label{sec:discussion}

\subsection{EPIC 201295312}
\label{sec:epic201295312}

EPIC 201295312b has an orbital period of 5.66 days. We do not clearly detect the RV amplitude caused by the planet, but do find a longer-period trend. Using the procedure described in Section~\ref{sec:orbital}, we find the long term drift to be adequately described by a second order polynomial, with a linear term of 3.3~cm~s$^{-1}$~d$^{-1}$ and a quadratic term of 1.83~cm~s$^{-1}$~d$^{-1}$, while using 2457099.654 HJD as the constant time of reference. 

\begin{figure}[ht]
\includegraphics[height=7cm]{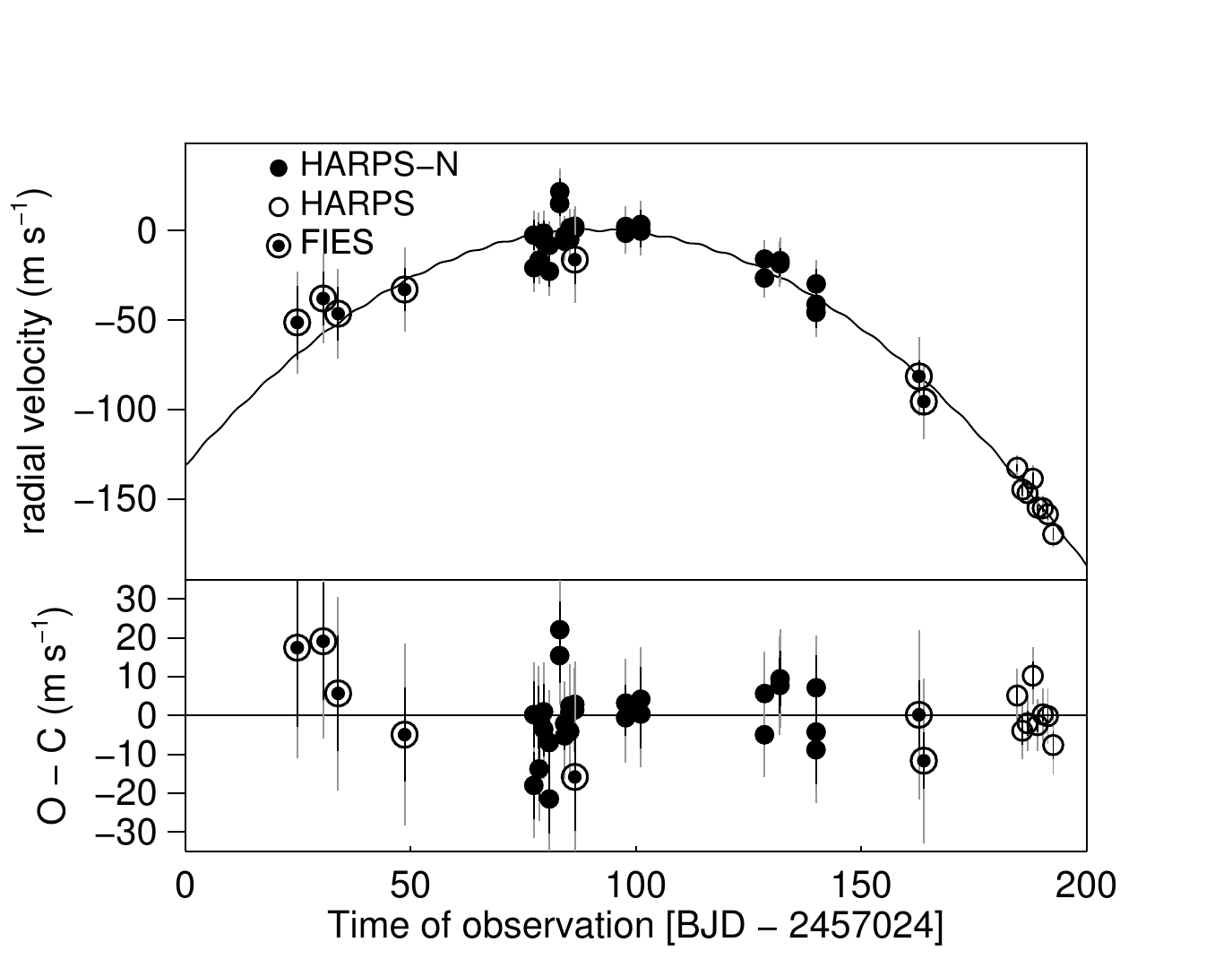}
\caption{{RV observations of EPIC 201295312 as function of time. The best fitted model using a quadratic long term trend is shown along with the data. Assuming good agreement between the velocity offsets of HARPS and HARPS-N (see also Section~\ref{sec:orbital}) the data does require a quadratic term to be adequately fitted.}\label{fig:epic201295312_rv_trend}}
\end{figure}

{We speculate that this trend, which is shown in Figure~\ref{fig:epic201295312_rv_trend}, could be caused by the gravitational influence of an additional body in the system, such that additional monitoring might reveal the orbital period and mass function of this presumed companion. Our currently data constrains it poorly, but as one example of what may be causing it, we find that a circular orbit with a period of 365 days leads to a good fit with a $K_\star$ amplitude of 155~m~s$^{-1}$. Assuming an inclination of 90 degrees, this implies a mass of 5.9~$M_\textrm{Jup}$. However, we caution against overinterpreting these values, as only a small part of such an orbit is covered with the current data, and longer orbital periods cannot be excluded.}

\begin{figure}[ht]
\centering
\resizebox{\hsize}{!}{\includegraphics{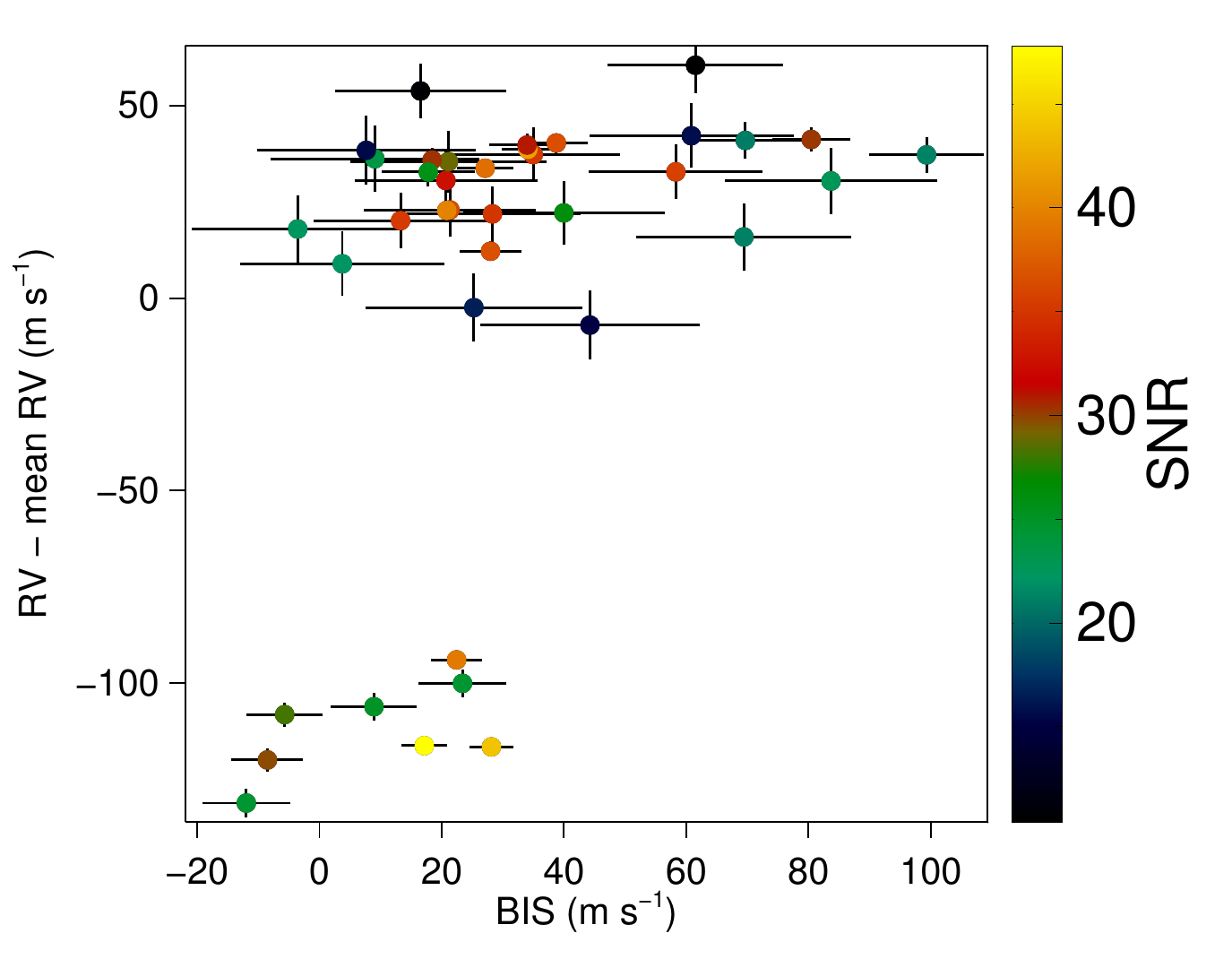}}
\caption{{BIS from HARPS and HARPS-N CCFs for EPIC 2012995312 plotted versus the stellar RVs. The color code indicates the signal-to-noise ratio in the stellar spectra obtained around a wavelength range of $5560$\,\AA. The BIS values for the low RV points appear slightly shifted from the RV points with higher values, which would indicate the presence of a self-luminous companion, but because those data are taken with different telescopes the results are inconclusive. The uncertainty in the bisector values is taken to be twice the uncertainty in the RV values.}}
\label{fig:bisector_201295312}
\end{figure}

{We furthermore tested if the bisector measurements give any indication that the observed RV trend is caused by a stellar companion. The results are inconclusive: the HARPS and HARPS-N CCFs appear to show a small difference, but it is possible this is caused by the atmospheric conditions under which the stars were observed, or by small differences between the two instruments. The data are shown in Figure~\ref{fig:bisector_201295312}.}

\begin{figure}
\includegraphics[height=7cm]{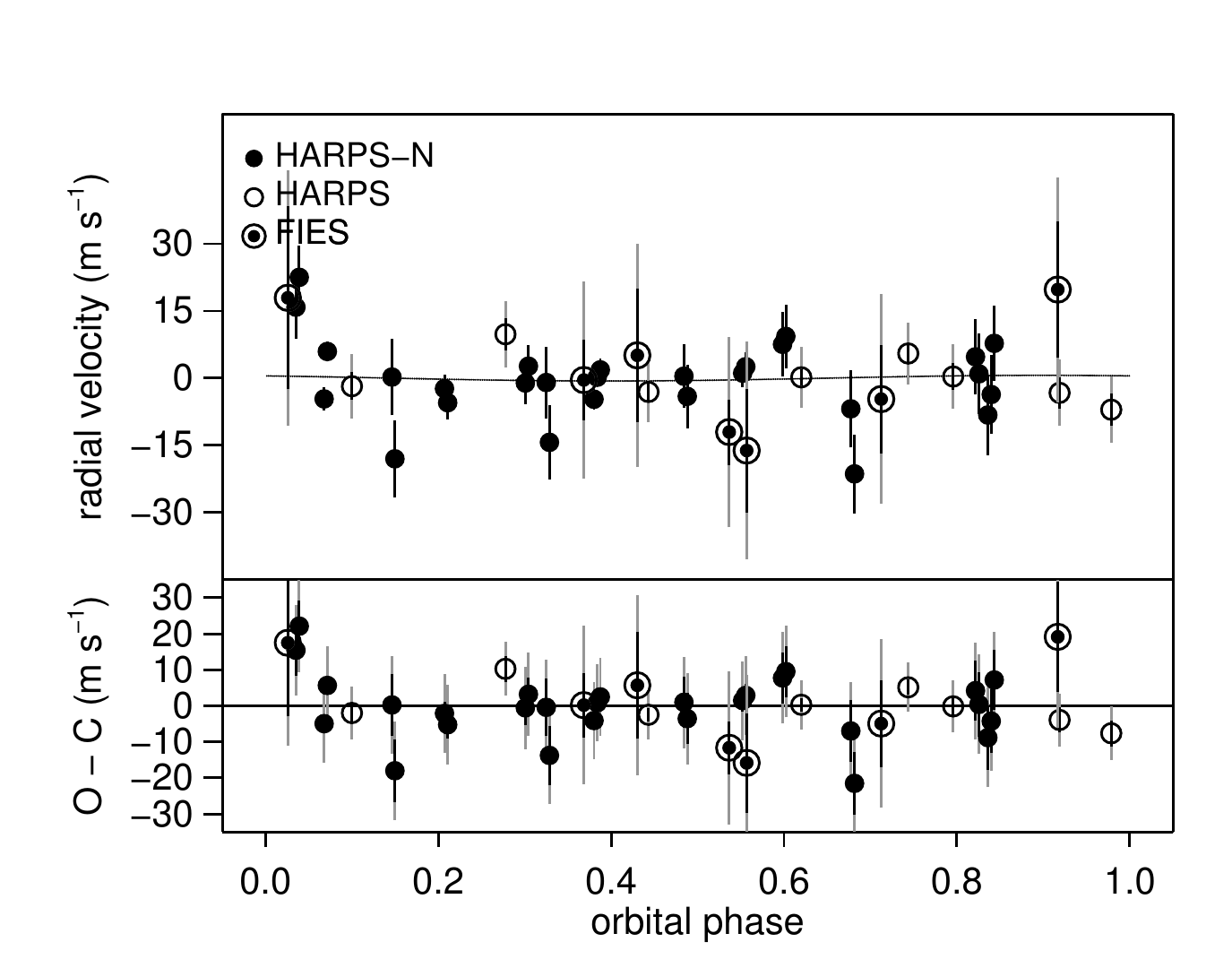}

\includegraphics[height=7cm]{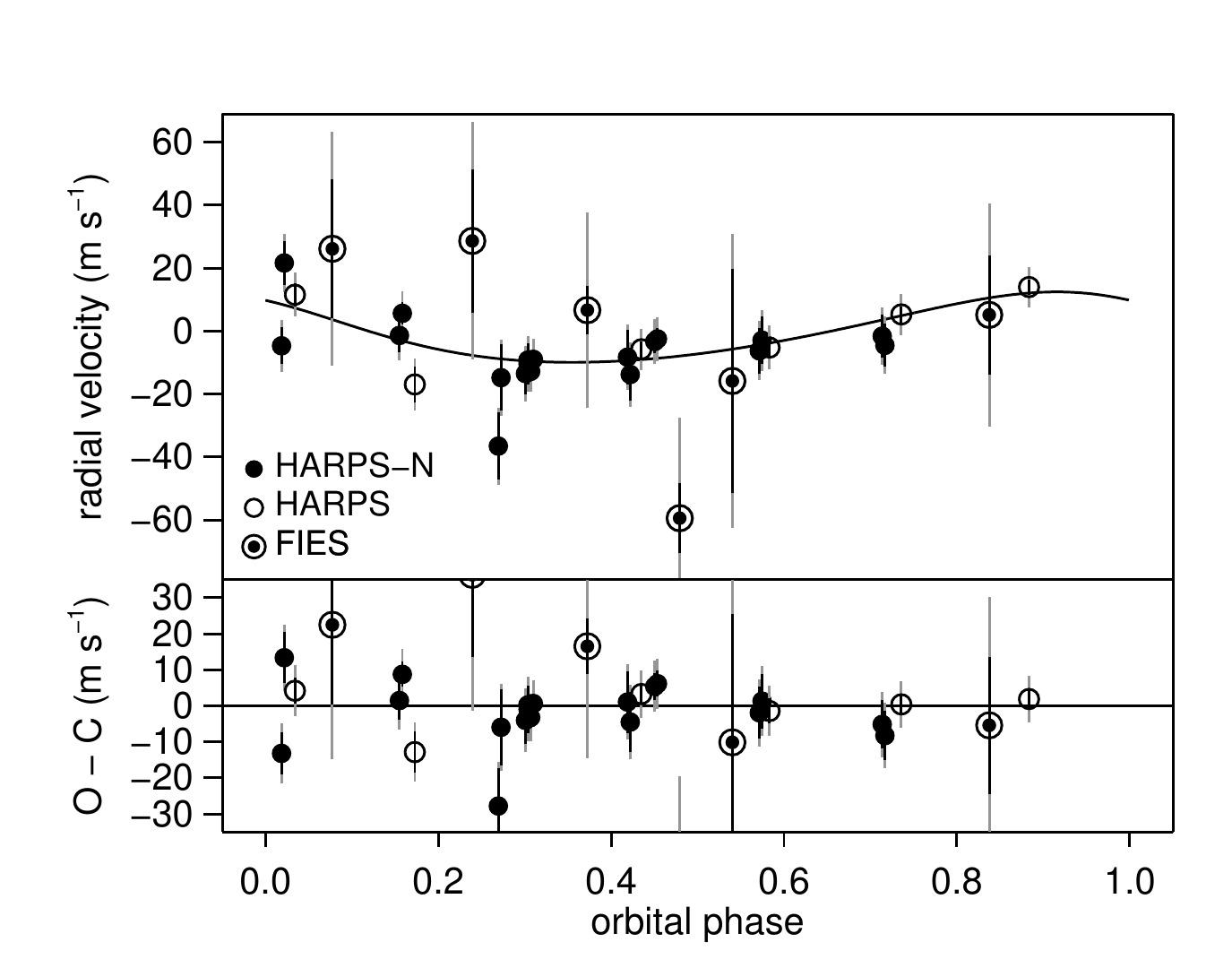}

\includegraphics[height=7cm]{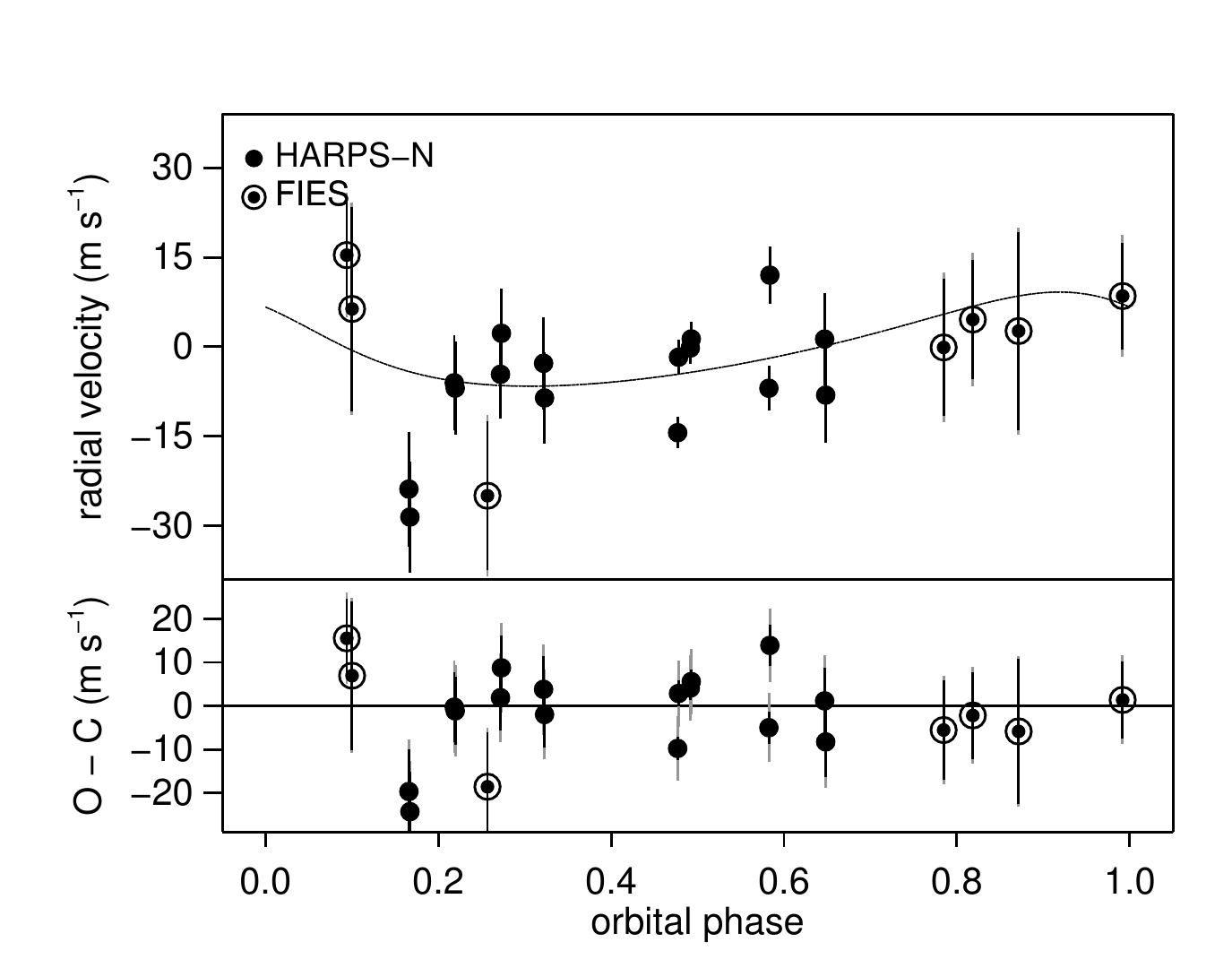}
\caption{RV observations over orbital phase for EPIC 201295312 (top), K2-27 (EPIC 201546283, middle) and K2-10 (EPIC 201577035, bottom). The best fitted model is shown with a solid line as well as the residuals after subtracting the model. The internal RV uncertainties are indicated by the black error bars, while the gray error bars include an additional ``stellar jitter'' term as explained in the text. Note that for K2-27 the residual plot do not show all FIES RVs due to the small RV interval displayed here. For this system the FIES data does not carry a lot of weight for the final solution and the zoom allows for a better inspection of the HARPS and HARPS-N residual which do determine our final solution.  \label{fig:rvs}}
\end{figure}

We {place an upper limit on the planetary mass of 12 $M_\oplus$}, which together with a measured radius of $2.75^{+0.24}_{-0.22}~\mathrm{R_\oplus}$, results in a planet density upper limit of 3.3 g~cm$^{-3}$. These limit are one-sided 95\% confidence intervals. We note that our best measured mass, $-0.5^{+7.6}_{-7.5} M_\oplus$, {the median of the distribution,} is negative (see Figure~\ref{fig:rvs}). We allow for negative (unphysical) mass to avoid positively biasing mass measurements for small planets. {While this can easily be avoided using a prior that prohibits the unphysical mass regime, we prefer not to do so because the negative masses are a statistically important measurement of the planetary mass \citep[see e.g.][for a detailed discussion]{marcy2014}. Allowing negative masses accounts naturally for the uncertainty in planet mass due to RV errors, and is key to allow unbiased constraints the interior structure based on a sample of small planets \citep[see e.g.][]{rogers2015,wolfgang2015}.} All 
parameters are available in Table~\ref{tab:planet_para}. 

\subsection{K2-27 (EPIC 201546283)}

\begin{figure}[!htb]
\centering
\resizebox{\hsize}{!}{\includegraphics{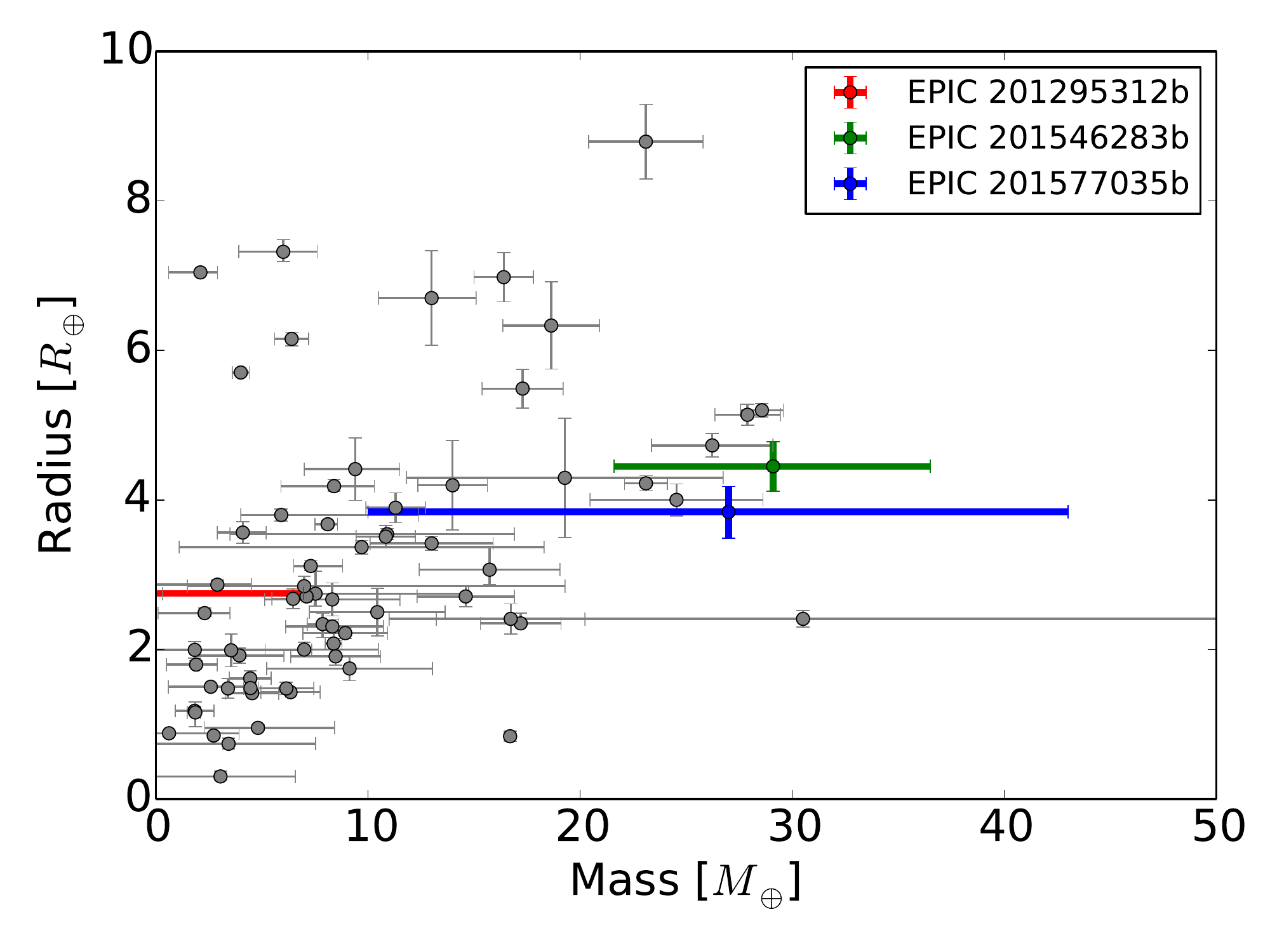}}
\caption{{Mass-radius diagram for transiting exoplanets smaller than $10~R_\oplus$ and less massive than $50~M_\oplus$ with masses measured from RV observations (data from exoplanets.org) and TTVs. Our new measurements are indicated.}
\label{fig:massradius}}
\end{figure}

\begin{figure}[ht]
\centering
\resizebox{\hsize}{!}{\includegraphics{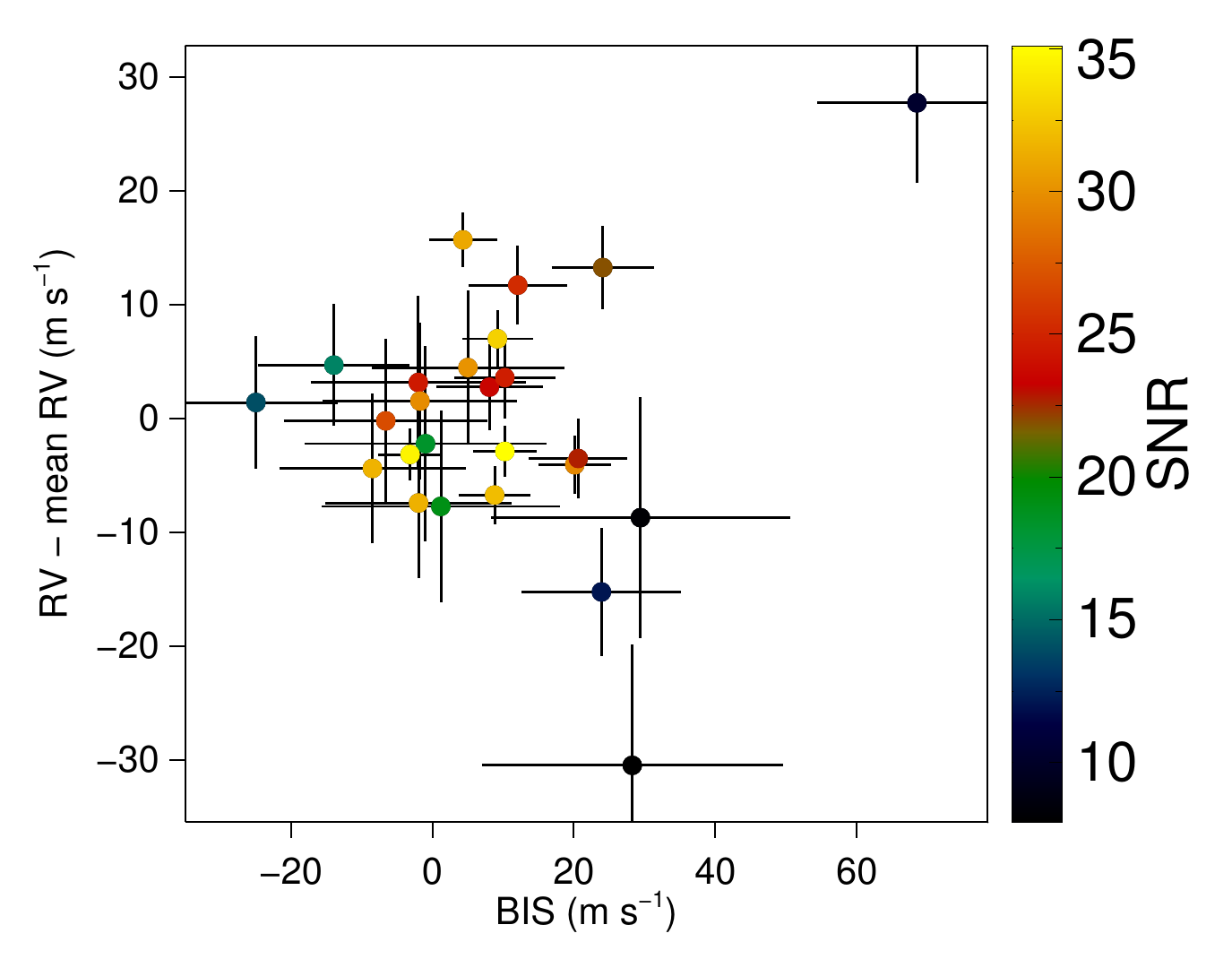}}
\caption{{BIS from HARPS and HARPS-N CCFs for K2-27 plotted versus the stellar RVs. The color code indicates the signal-to-noise ratio in the stellar spectra obtained around a wavelength range of $5560$\,\AA. There is no evidence for correlation between the BIS and RV. As expected lower signal-to-noise spectra do have a larger scatter in both BIS and RV. {The uncertainty in the bisector values is taken to be twice the uncertainty in the RV values.}}
\label{fig:bisectorplot}}
\end{figure}

For EPIC 201546283b, which has an orbital period $P$ of 6.77 days, we obtain a $3\sigma$ mass detection of $29.1^{+7.5}_{-7.4}~\mathrm{M_\oplus}$. This confirms the planetary nature of this candidate, which we further refer to as K2-27b. Earlier work detected the transits of this candidate but was unable to confirm the planetary nature on statistical grounds \citep{montet2015}. We find a planetary radius of $4.45^{+0.33}_{-0.33}~\mathrm{R_\oplus}$, which taken together with the mass measurement leads to a planet density of $1.80^{+0.70}_{-0.55}~\mathrm{g~cm^{-3}}$. This makes this planet rather similar to Neptune (which has a density of 1.64$~\mathrm{g~cm^{-3}}$). As for EPIC 201295312 we allowed for the presence of a linear drift but found that it did not significantly change the results, and therefore we set it to zero. All parameters for this star and its planet are available in Table~\ref{tab:planet_para}. The planet is plotted on a mass-radius diagram in Figure~\ref{fig:massradius}.
{We now check if the RV signal could be caused by stellar activity. For this we calculate the BIS as defined by \cite{queloz2001} from the HARPS and HARPS-N CCFs and searched for a correlation with the measured RVs. If such a correlation does exist, then this is a sign of a deformation of the stellar absorption lines by stellar activity instead of a Doppler shift of the CCF caused by the gravitational pull of an unseen companion. However, no such correlation can be found. {We calculate the Pearson correlation coefficient, which is 0.105. With 23 degrees of freedom, we also find a t-statistic of 0.505, and a two-tailored test leads to a p-value of 0.61. All these tests indicate there is no significant evidence for a correlation between the BIS and RVs. The values are shown in Figure~\ref{fig:bisectorplot}.}}

\subsection{K2-10 (EPIC 201577035)}

K2-10b (EPIC 201577035b) was previously validated to be a true planet by \cite{montet2015}. Here we refine the stellar and planetary parameters and {constrain} the planet's mass. The planet is smaller than Neptune with an orbital period of 19.3 days and a radius of $3.84^{+0.35}_{-0.34}~\mathrm{R_\oplus}$. We measure its mass to be $27^{+17}_{-16}~\mathrm{M_\oplus}$, resulting in a planetary density of $2.6^{+2.1}_{-1.6}~{\rm g~cm}^{-3}$. {Within 95\% confidence, the planetary mass is below $57~\mathrm{M_\oplus}$.} We found no evidence for a linear drift and set it to zero in the final fit. We note that, due to uncooperative weather, the coverage of this system shows a significant phase gap (see Figure~\ref{fig:rvs}). All parameters are listed in Table~\ref{tab:planet_para}, and the system is indicated on a mass-radius diagram in Figure~\ref{fig:massradius}.

\section{Discussion}

We have reported our results for three planet candidates observed with K2 during Campaign~1. These planets have been found in the K2 data using two different algorithms and were also discussed by \cite{montet2015}. We measured the mass for the largest of these planets (EPIC 201546283b) and obtain lower significance measurements or upper limits of the masses and densities of the other two systems (EPIC 201295312b and EPIC 201577035b). The first of these planets is similar to Neptune, while the other two are smaller than Neptune. {For EPIC 201295312, we also discover a long-period trend indicative of an additional body.} Relatively few mass measurements are available for sub-Neptunes due to the small RV amplitudes these planets cause. For example, an extensive Keck campaign {following up on 22 \textit{Kepler} stars with known transiting} planets recently lead to 16 secure mass detections \citep{marcy2014} {and 26 more marginal measurements or upper limits}. Despite thousands of planetary 
candidates found by the primary \textit{Kepler} mission, many of those are too faint for follow-up measurements. Here, {\it K2} has the potential to make a significant contribution.

{Finally, the significant effort required to measure masses for the planets highlights the need for future missions such as TESS \citep{ricker2014} and PLATO \citep{rauer2014} which will observe even brighter stars.}

\acknowledgements

{\small We appreciate the quick and thoughtful comments and suggestions by the referee, which significantly improved the manuscript. We are grateful for the Python implementation of a BLS algorithm by Ruth Angus and Dan Foreman-Mackey. We are thankful to the GAPS consortium, which kindly agreed to exchange time with us, and the KEST team which shared observations. Based on observations made with the NOT and TNG telescopes operated on the island of La Palma in the Observatorio del Roque de Los Muchachos of the Instituto de Astrof\'\i sica de Canarias, as well as observations with the HARPS spectrograph at ESO's La Silla observatory (095.C-0718(A)). Funding for the Stellar Astrophysics Centre is provided by The Danish National Research Foundation (Grant agreement no.: DNRF106). The research is supported by the ASTERISK project (ASTERoseismic Investigations with SONG and Kepler) funded by the European Research Council (Grant agreement no.: 267864). We acknowledge ASK for covering travels in relation to this 
publication. 
This research has made use of the Exoplanet Orbit Database and the Exoplanet Data Explorer at exoplanets.org. This work was performed [in part] under contract with the Jet Propulsion Laboratory (JPL) funded by NASA
through the Sagan Fellowship Program executed by the NASA Exoplanet Science Institute. I.R. and M.P. acknowledge support from the Spanish Ministry of Economy and Competitiveness (MINECO) and the Fondo Europeo de Desarrollo Regional (FEDER) through grants ESP2013-48391-C4-1-R and ESP2014-57495-C2-2-R. N.N. acknowledges supports by the NAOJ Fellowship, Inoue Science Research Award, and Grant-in-Aid for Scientific Research (A) (JSPS KAKENHI Grant Number 25247026). T.H. is supported by Japan Society for Promotion of Science (JSPS) Fellowship for Research (No.25-3183). L.A.R. gratefully acknowledges support provided by NASA through Hubble Fellowship grant \#HF-51313 awarded by the Space Telescope Science Institute, which is operated by the Association of Universities for Research in Astronomy, Inc., for NASA, under contract NAS~5-26555. HD acknowledges support by grant AYA2012-39346-C02-02 of the Spanish Secretary of State for R\&D\&i (MICINN).}

\bibliographystyle{bibstyle}

\end{document}